\documentclass[journal]{IEEEtran}

\pdfoutput=1

\usepackage{amsmath}
\usepackage{epsfig}
\usepackage{amsfonts}
\usepackage[hypertexnames=false]{hyperref}
\usepackage{algorithm}
\usepackage{algorithmic}
\usepackage{url}

\newcommand{\field}[1]{\mathbb{#1}}
\def\R{\field{R}}                                % Blackboard R - Reals
\newcommand{\Ascr}{{\cal A}}

\newcommand{\Fscr}{{\cal F}}
\newcommand{\Gscr}{{\cal G}}
\newcommand{\Xscr}{{\cal X}}

\newcommand\1{{\bf 1}}

\newtheorem{theorem} {Theorem}
\newtheorem{conjecture} {Conjecture}

\newtheorem{lemma} {Lemma}

\newcommand{\tends}{\rightarrow}

\newcommand\minimize{\mathop{\mbox{\rm minimize}}\limits}

\newcommand\diag{\mathop{\mbox{\rm diag}}}

\begin{document}

\bibliographystyle{IEEEtran}

\title{Consensus Propagation}

\author{Ciamac~C.~Moallemi and~Benjamin~Van~Roy,~\IEEEmembership{Member,~IEEE}%
\thanks{Manuscript recieved June 28, 2005; revised July 18 2006.
    The first author was supported by a Benchmark Stanford Graduate
    Fellowship. This research was supported in part by the National
    Science Foundation through grant IIS-0428868 and a supplement to
    grant ECS-9985229 provided by the Management of Knowledge
    Intensive Dynamic Systems Program (MKIDS). The material in this paper was presented in part at the Neural Information Processing Systems (NIPS) conference, Vancouver, Canada, December 2005.}%
\thanks{C. C. Moallemi is with the Department of Electrical
    Engineering, Stanford University, Stanford, CA 94305 USA \{e-mail: ciamac@stanford.edu\}.}%
\thanks{B. Van Roy is with the Department of of Management Science \& Engineering and the Department of Electrical
    Engineering, Stanford University, Stanford, CA 94305 USA \{e-mail: bvr@stanford.edu\}.}}

\markboth{IEEE Transactions on Information Theory,~Vol.~52, No.~11,~November~2006}{Moallemi \MakeLowercase{\textit{et al.}}: Consensus Propagation}

\pubid{0000--0000/00\$00.00~\copyright~2006 IEEE}

\maketitle

\begin{abstract}
We propose {\it consensus propagation}, an asynchronous distributed
protocol for averaging numbers across a network.  We establish
convergence, characterize the convergence rate for regular
graphs, and demonstrate that the protocol exhibits better scaling
properties than {\it pairwise averaging}, an alternative that
has received much recent attention.  Consensus propagation can be
viewed as a special case of belief propagation, and our results
contribute to the belief propagation literature.  In particular, beyond 
singly-connected graphs, there are very few classes of relevant problems 
for which belief propagation is known to converge.
\end{abstract}


\section{Introduction}

\PARstart{C}{onsider} a network of $n$ nodes in which the $i$th node observes a
real number $y_i \in \R$ and aims to compute the average
$\bar{y}=\sum_{i=1}^n y_i / n$.  The design of scalable distributed
protocols for this purpose has received much recent attention and is
motivated by a variety of potential needs.  In both wireless sensor
and peer-to-peer networks, for example, there is interest in simple
protocols for computing aggregate statistics (see, e.g.
\cite{Intanagonwiwat00,Madden02,Madden02b,
  Zhao03,Bawa03,Jelasity04,Montresor04}), and averaging enables
computation of several important ones.  Further, averaging serves as a
primitive in the design of more sophisticated distributed information
processing algorithms.  For example, a maximum likelihood estimate can
be produced by an averaging protocol if each node's observations are
linear in variables of interest and noise is Gaussian
\cite{Xiao05}. \cite{Xiao05b} considers an averaging problem with
applications to load balancing and clock synchronization.  As another
example, averaging protocols are central to policy-gradient-based
methods for distributed optimization of network performance
\cite{Moallemi04}.

In this paper we propose and analyze a new protocol -- {\bf consensus
  propagation} -- for distributed averaging.  The protocol can operate
asynchronously and requires only simple iterative computations at
individual nodes and communication of parsimonious messages between
neighbors.  There is no central hub that aggregates information.  Each
node only needs to be aware of its neighbors -- no further information
about the network topology is required.  There is no need for
construction of a specially-structured overlay network such as a
spanning tree.  It is worth discussing two previously proposed and
well-studied protocols that also exhibit these features:
\begin{enumerate}
\item {\bf (probabilistic counting)} This protocol is based on ideas
  from \cite{Flajolet85} for counting distinct elements of a database
  and in \cite{Considine04} was adapted to produce a protocol for
  averaging.  The outcome is random, with variance that becomes
  arbitrarily small as the number of nodes grows.  However, for
  moderate numbers of nodes, say tens of thousands, high variance
  makes the protocol impractical.  The protocol can be repeated in
  parallel and results combined in order to reduce variance, but this
  leads to onerous memory and communication requirements. Convergence
  time of the protocol is analyzed in \cite{MoskAoyama05}.

\item {\bf (pairwise averaging)} In this protocol, each node maintains
  its current estimate of the average, and each time a pair of nodes
  communicate, they revise their estimates to both take on the mean of
  their previous estimates.  Convergence of this protocol in a very 
  general model of asynchronous computation and
  communication was established in \cite{Tsitsiklis84}, and there has been
  significant follow-on work, a recent sample of which is \cite{Blondel05}.  
  Recent work \cite{Kempe04,Boyd05} has studied the convergence rate and its
  dependence on network topology and how pairs of nodes are sampled.
  Here, sampling is governed by a certain doubly stochastic matrix, and 
  the convergence rate is characterized by its second-largest eigenvalue.
\end{enumerate}

\pubidadjcol

In terms of convergence rate, probabilistic counting 
dominates both pairwise averaging and consensus propagation in the
asymptotic regime.  However, consensus propagation and
pairwise averaging are likely to be more effective in moderately-sized
networks (up to hundreds of thousands or perhaps even millions of nodes).
Further, these two protocols are both naturally studied as iterative
matrix algorithms.  As such, pairwise averaging will serve as a
baseline to which we will compare consensus propagation.

Consensus propagation is a simple algorithm with an intuitive
interpretation.  It can also be viewed as an asynchronous distributed
version of belief propagation as applied to approximation of
conditional distributions in a Gaussian Markov random field.  When the
network of interest is singly-connected, prior results about
belief propagation imply convergence of consensus propagation.
However, in most cases of interest, the network is not
singly-connected and prior results have little to say about
convergence.  In particular, Gaussian belief propagation on a 
graph with cycles is not guaranteed to converge, as demonstrated 
by numerical examples in \cite{Rusmevichientong01}.

In fact, there are very few relevant cases where belief propagation on
a graph with cycles is known to converge.  Some fairly general
sufficient conditions have been established
\cite{Tatikonda02,Heskes04,Ihler05,Mooij05}, but these conditions are
abstract and it is difficult to identify interesting classes of
problems that meet them.  One simple case where belief propagation is
guaranteed to converge is when the graph has only a single cycle and
variables have finite support \cite{Forney98,Aji98,Weiss00}.  In its
use for decoding low-density parity-check codes, though convergence
guarantees have not been made, \cite{Richardson01} establishes
desirable properties of iterates, which hold with high probability.
Recent work proposes the use of belief propagation to solve
maximum-weight matching problems and proves convergence in that
context \cite{bayati05}. In the Gaussian case,
\cite{Rusmevichientong01,Weiss01} provide sufficient conditions for
convergence, but these conditions are difficult to interpret and do
not capture situations that correspond to consensus propagation. Since
this paper was submitted for publication, a general class of results
has been developed for the convergence of Gaussian belief propagation
\cite{Johnson06,Moallemi06}. These results can be viewed as a
generalization of the convergence results in this paper. However, they
do not address the issue of rate of convergence.

With this background, let us discuss the primary contributions of this paper:
\begin{enumerate}
\item We propose consensus propagation, a new asynchronous distributed protocol
  for averaging.
\item We prove that consensus propagation converges even when executed
  asynchronously.  Since there are so few classes of relevant problems
  for which belief propagation is known to converge, even with {\it
    synchronous execution}, this is surprising. 
\item We characterize the convergence time in regular graphs of the
  synchronous version of consensus propagation in terms of the the
  mixing time of a certain Markov chain over edges of the graph.
\item We explain why the convergence time of consensus propagation scales
  more gracefully with the number of nodes than does that of pairwise averaging,
  and for certain classes of graphs, we quantify the improvement.
\end{enumerate}

It is worth mentioning a recent and related line of research on the
use of belief propagation as an asynchronous distributed protocol to
arrive at consensus among nodes in a network, when each node makes a
conditionally independent observation of the class of an object and
would like to know the most probable class based on all observations
\cite{Saligrama05}.  The authors establish that belief propagation
converges and provides each node with the most probable class when the
network is a tree or a regular graph.  They further show that for a
certain class of random graphs, the result holds in an asymptotic
sense as the number of nodes grows.  To deal with general connected
graphs, the authors offer a more complex protocol with convergence
guarantees.  It is interesting to note that this classification
problem can be reduced to one of averaging.  In particular, if each
node starts out with the conditional probability of each class given
its own observation and the network carries out a protocol to compute
the average log-probability for each class, each node obtains the
conditional probabilities given {\it all} observations.  Hence,
consensus propagation also solves this classification problem.

\section{Algorithm}

Consider a connected undirected graph $(V,E)$ with
$V=\{1,\ldots,n\}$. For each node $i \in V$, let $N(i) = \{ j\ |\
(i,j)\in E\}$ be the set of neighbors of $i$.  Let $\vec{E} \subseteq
V\times V$ be a set consisting of two directed edges $\{i,j\}$ and
$\{j,i\}$ per undirected edge $(i,j) \in E$.  (In general, we will use
braces for directed edges and parentheses for undirected edges.)

Each node $i\in V$ is assigned a number $y_i\in \R$.  The goal is for
each node to obtain an estimate of $\bar{y} = \sum_{i\in V} y_i/n$
through an asynchronous distributed protocol in which each node
carries out simple computations and communicates parsimonious messages
to its neighbors.

We propose consensus propagation as an approach to the aforementioned
problem.  In this protocol, if a node $i$ communicates to a neighbor
$j$ at time $t$, it transmits a message consisting of two numerical
values.  Let $\mu^{(t)}_{ij} \in \R$ and $K^{(t)}_{ij} \in \R_+$
denote the values associated with the most recently transmitted
message from $i$ to $j$ at or before time $t$.  At each time $t$, node
$i$ has stored in memory the most recent message from each neighbor:
$\{\mu^{(t)}_{ui}, K^{(t)}_{ui}\ |\ u \in N(i)\}$.  If, at time $t+1$,
node $i$ chooses to communicate with a neighboring node $j \in N(i)$,
it constructs a new message that is a function of the set of most
recent messages $\{ \mu^{(t)}_{ui}, K^{(t)}_{ui}\ |\ u \in
N(i)\setminus j\}$ received from neighbors other than $j$. The initial
values in memory before receiving any messages are arbitrary.

In order to illustrate how the parameter vectors $\mu^{(t)}$ and
$K^{(t)}$ evolve, we will first describe a special case of the
consensus propagation algorithm that is particularly intuitive. Then,
we will describe the general algorithm and its relationship to belief
propagation.

\subsection{Intuitive Interpretation}\label{se:intuitive}

\begin{figure*}[htb]
\centering
\includegraphics{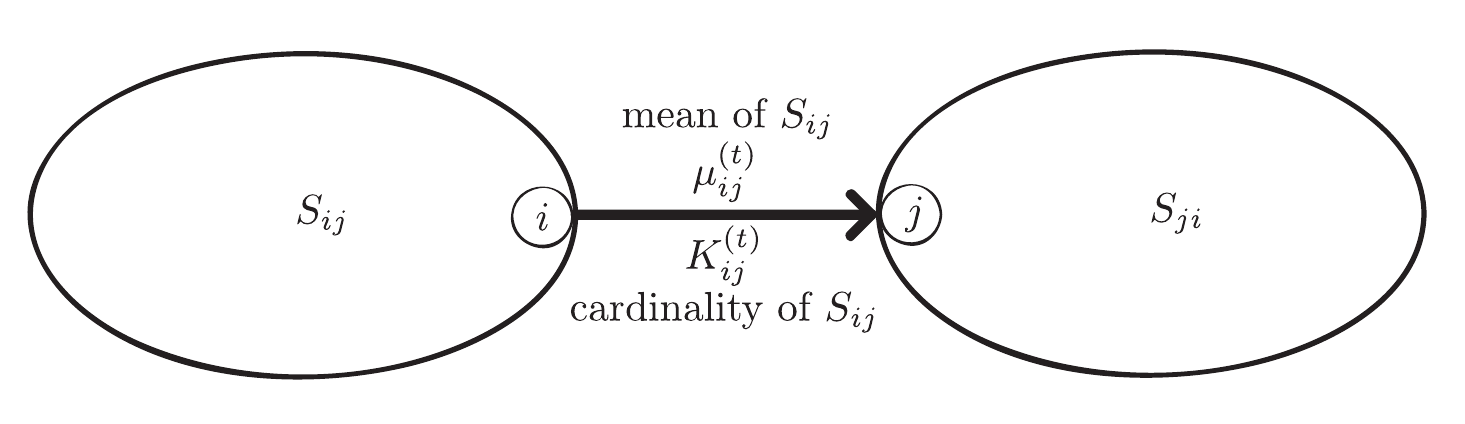}
\caption{Interpretation of messages in a singly connected graph with $\beta = \infty$.\label{fig:interpretation}}
\end{figure*}

Consider the special case of a singly-connected graph. That is, a
connected graph where there are no loops present (a tree). Assume, for
the moment, that at every point in time, every pair of connected nodes
communicates. As illustrated in Fig.~\ref{fig:interpretation}, for
any edge $\{i,j\}\in \vec{E}$, there is a set $S_{ij} \subset V$ of
nodes, with $i \in S_{ij}$, that can transmit information to $S_{ji} =
V \setminus S_{ij}$, with $j \in S_{ji}$, only through $\{i,j\}$.  In
order for nodes in $S_{ji}$ to compute $\overline{y}$, they must at
least be provided with the average $\mu^*_{ij}$ among observations at
nodes in $S_{ij}$ and the cardinality $K^*_{ij} =
|S_{ij}|$. Similarly, in order for nodes in $S_{ij}$ to compute
$\overline{y}$, they must at least be provided with the average
$\mu^*_{ji}$ among observations at nodes in $S_{ji}$ and the
cardinality $K^*_{ji}=|S_{ji}|$. These values must be communicated
through the link $\{j,i\}$.

The messages $\mu^{(t)}_{ij}$ and $K^{(t)}_{ij}$, transmitted from
node $i$ to node $j$, can be viewed as iterative 
estimates of the quantities $\mu^*_{ij}$ and $K^*_{ij}$.  They evolve
according to
\begin{subequations}
\begin{align}
\mu^{(t)}_{ij} & = 
\frac{y_i + \sum_{u \in N(i) \setminus j} K^{(t-1)}_{ui} \mu^{(t-1)}_{ui}}
{1 + \sum_{u \in N(i) \setminus j} K^{(t-1)}_{ui}},
&\forall\ \{i,j\}\in\vec{E},
\\
K^{(t)}_{ij} & = 1 + \sum_{u \in N(i) \setminus j} K^{(t-1)}_{ui},
&\forall\ \{i,j\}\in\vec{E}.
\label{eq:K-infty}
\end{align}
\end{subequations}
At each time $t$, each node $i$ computes an estimate of the global
average $\bar{y}$ according to
\[
x^{(t)}_i = \frac{y_i + \sum_{u \in N(i)} K^{(t)}_{ui} \mu^{(t)}_{ui}}
{1 + \sum_{u \in N(i)} K^{(t)}_{ui}}.
\]

Assume that the algorithm is initialized with $K^{(0)}=0$. A simple
inductive argument shows that at each time $t \geq 1$,
$\mu_{ij}^{(t)}$ is the average among observations at the nodes in the
set $S_{ij}$ that are at a distance less than or equal to $t$ from
node $i$. Furthermore, $K^{(t)}_{ij}$ is the cardinality of this
collection of nodes. Since any node in $S_{ij}$ is at a distance from
node $i$ that it at most the diameter of the graph, if $t$ is greater
that the diameter of the graph, we have $K^{(t)} = K^*$ and $\mu^{(t)}
= \mu^*$.  Thus, for any $i \in V$, and $t$ sufficiently large,
\[
x^{(t)}_i = \frac{y_i + \sum_{u \in N(i)} K^*_{ui} \mu^*_{ui}}
{1 + \sum_{u \in N(i)} K^*_{ui}} = \overline{y}.
\]
So, $x_i^{(t)}$ converges to the global average $\overline{y}$.
Further, this simple algorithm converges in as short a time as is
possible, since the diameter of the graph is the minimum amount of
time for the two most distance nodes to communicate.

Now, suppose that the graph has cycles.  For any directed edge
$\{i,j\} \in \vec{E}$ that is part of a cycle, $K^{(t)}_{ij}
\rightarrow \infty$. Hence, the algorithm does not converge.  A
heuristic fix might be to compose the iteration \eqref{eq:K-infty}
with one that attenuates:
\begin{align*}
\tilde{K}^{(t)}_{ij} & \leftarrow 1 + \sum_{u \in N(i) \setminus j} K^{(t-1)}_{ui},
\\
K^{(t)}_{ij} & \leftarrow \frac{\tilde{K}^{(t)}_{ij}}{1 +
\tilde{K}^{(t)}_{ij} / (\beta Q_{ij})}.
\end{align*}
Here, $Q_{ij} > 0$ and $\beta>0$ are positive constants. We can view
the unattenuated algorithm as setting $\beta=\infty$.  In the
attenuated algorithm, the message is essentially unaffected when
$\tilde{K}^{(t)}_{ij}/(\beta Q_{ij})$ is small but becomes
increasingly attenuated as $\tilde{K}^{(t)}_{ij}$ grows.  This is
exactly the kind of attenuation carried out by consensus propagation.
Understanding why this kind of attenuation leads to desirable results
is a subject of our analysis.

\subsection{General Algorithm}

Consensus propagation is parameterized by a scalar $\beta > 0$ and a
non-negative matrix $Q \in \R_+^{n\times n}$ with $Q_{ij} > 0$ if and
only if $i \neq j$ and $(i,j)\in E$.  For each $\{i,j\}
\in \vec{E}$, it is useful to define the following three functions:
\begin{subequations}
\begin{align}
\Fscr_{ij}(K) & = \frac{1 + \sum_{u \in N(i) \setminus j} K_{ui}}
{1 + \frac{1}{\beta Q_{ij}} 
\left(1 + \sum_{u \in N(i) \setminus j} K_{ui}\right)},
\label{eq:fdef}
\\
\Gscr_{ij}(\mu,K) & = \frac{y_i + \sum_{u \in N(i) \setminus j} K_{ui} \mu_{ui}}
{1 + \sum_{u \in N(i) \setminus j} K_{ui}},
\label{eq:gdef}
\\
\Xscr_i(\mu,K) & = \frac{y_i + \sum_{u \in N(i)} K_{ui} \mu_{ui}}
{1 + \sum_{u \in N(i)} K_{ui}}.
\label{eq:xdef}
\end{align}
\end{subequations}
For each $t$, denote by $U_t \subseteq \vec{E}$ the set of directed 
edges along which messages are transmitted at time $t$.  
Consensus propagation is presented below as Algorithm~\ref{alg:cp}.

\begin{algorithm}
\caption{Consensus propagation.}
\label{alg:cp}
\begin{algorithmic}[1]
\FOR{time $t=1$ to $\infty$}
\FORALL{$\{i,j\} \in U_t$}
\STATE $K_{ij}^{(t)} \leftarrow \Fscr_{ij}(K^{(t-1)})$
\STATE $\mu_{ij}^{(t)} \leftarrow \Gscr_{ij}(\mu^{(t-1)},K^{(t-1)})$
\ENDFOR
\FORALL{$\{i,j\} \notin U_t$}
\STATE $K_{ij}^{(t)} \leftarrow K_{ij}^{(t-1)}$
\STATE $\mu_{ij}^{(t)} \leftarrow \mu_{ij}^{(t-1)}$
\ENDFOR
\STATE $x^{(t)} \leftarrow \Xscr(\mu^{(t)},K^{(t)})$
\ENDFOR
\end{algorithmic}
\end{algorithm}

Consensus propagation is a {\it distributed protocol} because
computations at each node require only information that is locally
available.  In particular, the messages
$K^{(t)}_{ij}=\Fscr_{ij}(K^{(t-1)})$ and
$\mu^{(t)}_{ij}=\Gscr_{ij}(\mu^{(t-1)},K^{(t-1)})$ transmitted from
node $i$ to node $j$ depend only on $\{\mu^{(t-1)}_{ui},
K^{(t-1)}_{ui}\ |\ u \in N(i)\}$, which node $i$ has stored in memory.
Similarly, $x_i^{(t)}$, which serves as an estimate of $\overline{y}$,
depends only on $\{\mu^{(t)}_{ui}, K^{(t)}_{ui}\ |\ u \in N(i)\}$.

Consensus propagation is an {\it asynchronous protocol} because only a
subset of the potential messages are transmitted at each time.  Our
convergence analysis can also be extended to accommodate more general
models of asynchronism that involve communication delays, as those
presented in \cite{BertsekasPDP}.

In our study of convergence {\it time}, we will focus on the {\it
  synchronous} version of consensus propagation.  This is where $U_t =
\vec{E}$ for all $t$.  Note that synchronous consensus propagation is
defined by:
\begin{subequations}\label{eq:syncdef}
\begin{align}
K^{(t)} & = \Fscr(K^{(t-1)}),
\\
\mu^{(t)} & = \Gscr(\mu^{(t-1)},K^{(t-1)}), 
\\
x^{(t)} & = \Xscr(\mu^{(t-1)},K^{(t-1)}).
\end{align}
\end{subequations}

\subsection{Relation to Belief Propagation}

Consensus propagation can also be viewed as a special case of belief
propagation.  In this context, belief propagation is used to
approximate the marginal distributions of a vector $x \in \R^n$
conditioned on the observations $y \in \R^n$.  The mode of each of
the marginal distributions approximates $\overline{y}$.

Take the prior distribution over $(x,y)$ to be the normalized product
of potential functions $\{\psi_i(\cdot)\ |\ i \in V\}$ and compatibility
functions $\{ \psi^\beta_{ij}(\cdot)\ |\ (i,j) \in E\}$, given by
\begin{align*}
\psi_i(x_i) & = \exp(-(x_i-y_i)^2),
\\
\psi^\beta_{ij}(x_i,x_j) & = \exp(-\beta Q_{ij} (x_i-x_j)^2),
\end{align*}
where $Q_{ij}>0$, for each edge $(i,j) \in E$, and $\beta>0$ are constants.
Note that $\beta$ can be viewed as an inverse temperature parameter;
as $\beta$ increases, components of $x$ associated with adjacent nodes
become increasingly correlated.

Let $\Gamma$ be a positive semidefinite symmetric matrix such that
\[
x^\top \Gamma x = 
\sum_{(i,j)\in E} 
Q_{ij} (x_i - x_j)^2.
\]
Note that when $Q_{ij}=1$, for all edges $(i,j)\in E$, $\Gamma$ is the
graph Laplacian.  Given the vector $y$ of observations, the
conditional density of $x$ is
\[
\begin{split}
p^\beta(x) & \propto
\prod_{i\in V} \psi_i(x_i) \prod_{(i,j)\in E} \psi^\beta_{ij}(x_i,x_j)
\\
& = \exp\left( - \| x - y \|^2_2 - \beta x^\top\Gamma x \right).
\end{split}
\]

Let $x^\beta$ denote the mode (maximizer) of $p^\beta(\cdot)$.  Since the
distribution is Gaussian, each component $x_i^\beta$ is also the mode
of the corresponding marginal distribution.  Note that $x^\beta$ it is
the unique solution to the positive definite quadratic program
\begin{equation}\label{eq:opt}
\minimize_{x}\quad 
\| x - y \|^2_2 + \beta x^\top\Gamma x.
\end{equation}
The following theorem relates $x^\beta$ to the mean value $\bar{y}$.
\begin{theorem}\label{th:xstarconv}
$\sum_i x^\beta_i/n = \bar{y}$ and $\lim_{\beta \uparrow \infty}
  x^\beta_i = \bar{y}$, for all $i\in V$.
\end{theorem}
\begin{proof}
The first order conditions for optimality imply $(I + \beta
\Gamma)x^\beta = y$.  If we set $\1=(1,\ldots,1)^\top\in\R^n$, we have
$\Gamma \1 = 0$, hence $\1^\top x^\beta/n=\1^\top y/n=\bar{y}$. Let $U$ be
an orthogonal matrix and $D$ a diagonal matrix that form a spectral
decomposition of $\Gamma$, that is $\Gamma = U^\top D U$. Then, we have
$x^\beta = U^\top (I + \beta D)^{-1} U y$. It is clear that $\Gamma$ has
eigenvalue 0 with multiplicity 1 and corresponding normalized
eigenvector $\1/\sqrt{n}$, and all other eigenvalues $d_2,\ldots,d_n$
of $\Gamma$ are positive. Then, if $D=\diag(0,d_2,\ldots,d_n)$,
\[
\begin{split}
\lefteqn{\lim_{\beta\tends\infty} x^\beta} &
\\
& =
\lim_{\beta\tends\infty} 
U^\top \diag(1, 1/(1 + \beta d_2),\ldots,1/(1+\beta d_n)) U y
\\
& = \1\1^\top y/ n.
\end{split}
\]
\end{proof}
The above theorem suggests that if $\beta$ is sufficiently large, then
each component $x^\beta_i$ can be used as an estimate of $\bar{y}$.

In belief propagation, messages are passed along edges of a Markov
random field. In our case, because of the structure of the
distribution $p^\beta(\cdot)$, the relevant Markov random field has
the same topology as the graph $(V,E)$.  The message $M^{(t)}_{ij}(\cdot)$
passed from node $i$ to node $j$ at time $t$ is a distribution on the
variable $x_j$. Node $i$ computes this message using incoming messages
from other nodes as defined by the update equation
\begin{equation}\label{eq:bpupdate}
M^{(t)}_{ij}(x_j) = \kappa \int 
\psi_{ij}(x'_i,x_j)
\psi_{i}(x'_i)
\prod_{u\in N(i)\setminus j}
M^{(t-1)}_{ui}(x'_i)
\ 
dx'_i.
\end{equation}
Here, $\kappa$ is a normalizing constant.  Since our underlying
distribution $p^\beta(\cdot)$ is Gaussian, it is natural to consider
messages which are Gaussian distributions. In particular, let
$(\mu^{(t)}_{ij},K^{(t)}_{ij}) \in \R\times\R_+$ parameterize Gaussian message
$M^{(t)}_{ij}(\cdot)$ according to
\[
M^{(t)}_{ij}(x_j) \propto 
\exp\left(- K^{(t)}_{ij}(x_j-\mu^{(t)}_{ij})^2 \right).
\]
Then, \eqref{eq:bpupdate} is equivalent to the synchronous consensus
propagation iterations for $K^{(t)}$ and $\mu^{(t)}$.

The sequence of densities
\[
\begin{split}
p^{(t)}_j(x_j) 
&
\propto
\psi_j(x_j)
\prod_{i\in N(j)}
M^{(t)}_{ij}(x_j)
\\
&
=
\exp\left(
-(x_j - y_j)^2
-
\sum_{i\in N(j)}
K^{(t)}_{ij}(x_j - \mu^{(t)}_{ij})^2
\right),
\end{split}
\]
is meant to converge to an approximation of the marginal conditional
distribution of $x_j$.  As such, an approximation to $x^\beta_j$ is
given by maximizing $p^{(t)}_j(\cdot)$.  It is easy to show that, the
maximum is attained by $x^{(t)}_j = \Xscr_j(\mu^{(t)}, K^{(t)})$.
With this and aforementioned correspondences, we have shown that
consensus propagation is a special case of belief propagation, and
more specifically, Gaussian belief propagation.

Readers familiar with belief propagation will notice that in the
derivation above we have used the sum-product form of the algorithm.
In this case, since the underlying distribution is Gaussian, the
max-product form yields equivalent iterations.

\subsection{Relation to Prior Results}

In light of the fact that consensus propagation is a special case of
Gaussian belief propagation, it is natural to ask what prior results
on belief propagation --- Gaussian or more broadly --- have to say in
this context.  Results from
\cite{Weiss01,Rusmevichientong01,Wainwright03} establish that, in the
absence of degeneracy, Gaussian belief propagation has a unique fixed
point and that the mode of this fixed point is unbiased. The issue of
convergence, however, is largely poorly understood.  As observed
numerically in \cite{Rusmevichientong01}, Gaussian belief propagation
can diverge, even in the absence of degeneracy.  Abstract sufficient
conditions for convergence that have been developed in
\cite{Weiss01,Rusmevichientong01} are difficult to verify in the
consensus propagation case.

\section{Convergence}

As we have discussed, Gaussian belief propagation can diverge, even when the 
graph has a single cycle.  One might expect the same from consensus propagation.
However, the following theorem establishes convergence.

\begin{theorem}\label{th:conv}
The following hold:
\begin{itemize}{\setlabelwidth{(iii)}}
\item[(i)] There exist unique vectors $(\mu^\beta,K^\beta)$ such that
$K^\beta = \Fscr(K^\beta)$ and $\mu^\beta = \Gscr(\mu^\beta,K^\beta)$.
\item[(ii)] Suppose that each directed edge $\{i,j\}$
  appears infinitely often in the sequence of communication sets
  $\{U_t\}$. Then, independent of the initial condition $(\mu^{(0)},K^{(0)})$,
\[
\lim_{t\tends\infty} K^{(t)} = K^\beta,
\quad\text{and}\quad
\lim_{t\tends\infty} \mu^{(t)} = \mu^\beta.
\]
\item[(iii)] Given $(\mu^\beta,K^\beta)$, if
  $x^\beta=\Xscr(\mu^\beta,K^\beta)$, then $x^\beta$ is the mode of
  the distribution $p^\beta(\cdot)$.
\end{itemize}
\end{theorem}

Note that the condition on the communication sets in
Theorem~\ref{th:conv}(ii) corresponds to {\it total asynchronism} in
the language of \cite{BertsekasPDP}. This is a weak assumption which
ensures only that every component of $\mu^{(t)}$ and $K^{(t)}$ is updated
infinitely often.

The proof of this theorem is deferred until the appendix, but it rests
on two ideas. First, notice that, according to the update equation
\eqref{eq:fdef}, $K^{(t)}$ evolves independently of $\mu^{(t)}$. Hence, we
analyze $K^{(t)}$ first. Following the work in \cite{Rusmevichientong01},
we prove that the functions $\{\Fscr_{ij}(\cdot)\}$ are monotonic.
This property is used to establish convergence to a unique fixed
point. Next, we analyze $\mu^{(t)}$ assuming that $K^{(t)}$ has already
converged. Given fixed $K$, the update equations for $\mu^{(t)}$ are
linear, and we establish that they induce a contraction with respect
to the maximum norm. This allows us to establish existence of a fixed
point and both synchronous and asynchronous convergence.

\section{Convergence Time for Regular Graphs}\label{se:convrate}

In this section, we will study the convergence time of 
synchronous consensus propagation.  For $\epsilon > 0$, we will
say that an estimate $\tilde{x}$ of $\bar{y}$ is
$\epsilon$-accurate if
\begin{equation}\label{eq:eaccuratedef}
\| \tilde{x} - \bar{y} \1 \|_{2,n} \leq \epsilon.
\end{equation}
Here, for integer $m$, we set $\|\cdot\|_{2,m}$ to be the norm on
$\R^m$ defined by $\|x\|_{2,m}=\|x\|_2/\sqrt{m}$. We are interested in
the number of iterations required to obtain an $\epsilon$-accurate
estimate of the mean $\bar{y}$.

Note that we are primarily interested in how the performance of
consensus propagation behaves over a series of problem instances as we
scale the size of the graph. Since our measure of error
\eqref{eq:eaccuratedef} is absolute, we require that the set of values
$\{ y_i \}$ lie in some bounded set. Without loss of generality, we
will take $y_i \in [0,1]$, for all $i \in V$.

\subsection{The Case of Regular Graphs}

We will restrict our analysis of convergence time to cases where
$(V,E)$ is a $d$-regular graph, for $d \geq 2$.  Extension of our
analysis to broader classes of graphs remains an open issue.  We will
also make simplifying assumptions that $Q_{ij}=1$, $\mu^{(0)}_{ij} = y_i$,
and $K^{(0)} = [k_0]_{ij}$ for some scalar $k_0 \geq 0$.

In this restricted setting, the subspace of constant $K$ vectors is
invariant under $\Fscr$. This implies that there is some
scalar $k^\beta>0$ so that $K^\beta = [k^\beta]_{ij}$.
This $k^\beta$ is the unique solution to the fixed point equation
\begin{equation}\label{eq:kfixedpt}
k^\beta = \frac{1 + (d - 1) k^\beta}{1 + (1 + (d - 1) k^\beta)/\beta}.
\end{equation}
Given a uniform initial condition $K^{(0)} = [ k_0 ]_{ij}$, we can study
the sequence of iterates $\{ K^{(t)} \}$ by examining the scalar sequence
$\{ k_t \}$, defined by
\begin{equation}\label{eq:ktevolve}
k_t = \frac{1 + (d - 1) k_{t-1}}{1 + (1 + (d - 1) k_{t-1})/\beta}.
\end{equation}
In particular, we have $K^{(t)} = [k_t]_{ij}$, for all $t \geq 0$.

Similarly, in this setting, the equations for the evolution of $\mu^{(t)}$
take the special form
\[
\begin{split}
\mu^{(t)}_{ij}
& =
 \frac{y_i}{1 + (d-1)k_{t-1}}
\\
&
\quad
+\: \left(1 - \frac{1}{1 + (d-1)k_{t-1}}\right)
\sum_{u\in N(i)\setminus j} \frac{\mu^{(t-1)}_{ui}}{d-1}.
\end{split}
\]
Defining $\gamma_t=1/(1+(d-1)k_t)$, we have, in vector form,
\begin{equation}\label{eq:muiter}
\mu^{(t)} = \gamma_{t-1} \hat{y} + (1 - \gamma_{t-1}) \hat{P} \mu^{(t-1)},
\end{equation}
where $\hat{y}\in\R^{nd}$ is a vector with $\hat{y}_{ij} = y_i$ and
$\hat{P}\in\R_+^{nd\times nd}$ is a doubly stochastic matrix. The
matrix $\hat{P}$ corresponds to a Markov chain on the set of directed
edges $\vec{E}$.  In this chain, a directed edge $\{i,j\}$ transitions to a
directed edge $\{u,i\}$ with $u \in N(i) \setminus j$, with equal probability
assigned to each such edge. 
As in \eqref{eq:syncdef}, we associate each
$\mu^{(t)}$ with an estimate $x^{(t)}$ of $x^\beta$ according to
\[
x^{(t)} = \frac{1}{1 + d k^\beta} y + \frac{d k^\beta}{1 + d k^\beta} A \mu^{(t)},
\]
where $A\in \R_+^{n \times nd}$ is a matrix defined by 
$(A \mu)_j = \sum_{i \in N(j)} \mu_{ij}/d$.

\subsection{The Ces\`aro Mixing Time}

The update equation \eqref{eq:muiter} suggests that the convergence of
$\mu^{(t)}$ is intimately tied to a notion of mixing time associated with
$\hat{P}$.  Let $\hat{P}^\star$ be the Ces\`aro limit
\[
\hat{P}^\star =
\lim_{t\tends\infty} \sum_{\tau=0}^{t-1} \hat{P}^\tau/t.
\]
Define the {\it Ces\`aro mixing time} $\tau^\star$ by
\[
\tau^\star = 
\sup_{t\geq 0}
\left\| \sum_{\tau=0}^t (\hat{P}^\tau   - \hat{P}^\star )\right\|_{2,nd}.
\]
Here, $\|\cdot\|_{2,nd}$ is the matrix norm induced by the
corresponding vector norm $\|\cdot\|_{2,nd}$.  Since $\hat{P}$ is a
stochastic matrix, $\hat{P}^\star$ is well-defined and $\tau^\star <
\infty$. Note that, in the case where $\hat{P}$ is aperiodic,
irreducible, and symmetric, $\tau^\star$ corresponds to the
traditional definition of mixing time: the inverse of the spectral gap
of $\hat{P}$.

\subsection{Bounds on the Convergence Time}

Let $\gamma^\beta = \lim_{t \uparrow \infty} \gamma_t = 1 / (1 + (d-1)
k^\beta)$.  With an initial condition $k_0=k^\beta$, the update equation for
$\mu^{(t)}$ becomes
\[
\mu^{(t)} = \gamma^\beta \hat{y} + (1 - \gamma^\beta) \hat{P} \mu^{(t-1)}.
\]
Since $\gamma^\beta \in (0,1)$, this iteration is a contraction
mapping, with contraction factor $1-\gamma^\beta$.  It is easy to show
that $\gamma^\beta$ is monotonically decreasing in $\beta$, and as
such, large values of $\beta$ are likely to result in slower
convergence.  On the other hand, Theorem~\ref{th:xstarconv} suggests
that large values of $\beta$ are required to obtain accurate estimates
of $\bar{y}$.  To balance these conflicting issues, $\beta$ must be
appropriately chosen.

A time $t^*$ is said to be an {\it $\epsilon$-convergence time} if
estimates $x^{(t)}$ are $\epsilon$-accurate for all $t \geq t^*$.  The
following theorem, whose proof is deferred until the appendix,
establishes a bound on the $\epsilon$-convergence time of synchronous
consensus propagation given appropriately chosen $\beta$, as a
function of $\epsilon$ and $\tau^\star$.

\begin{theorem}\label{th:convrate}
Suppose $k_0 \leq k^\beta$.  If $d = 2$ there exists a
$\beta = \Theta((\tau^\star/\epsilon)^2)$ and 
if $d > 2$ there exists a $\beta = \Theta(\tau^\star/\epsilon)$
such that some $t^* = O((\tau^\star/\epsilon) \log(\tau^\star /\epsilon))$
is an $\epsilon$-convergence time.
\end{theorem}

In the above theorem, $k_0$ is initialized arbitrarily so long as
$k_0\leq k^\beta$. Typically, one might set $k_0=0$ to guarantee
this. Another case of particular interest is when $k_0=k^\beta$,
so that $k_t = k^\beta$ for all $t \geq 0$.  In this case, the following
theorem, whose proof is deferred until the appendix, offers a better
convergence time bound than Theorem~\ref{th:convrate}.

\begin{theorem}\label{th:preconvrate}
Suppose $k_0 = k^\beta$.  If $d = 2$ there exists a
$\beta = \Theta((\tau^\star/\epsilon)^2)$ and 
if $d > 2$ there exists a $\beta = \Theta(\tau^\star/\epsilon)$
such that some
$t^* = O((\tau^\star /\epsilon) \log(1/\epsilon))$
is an $\epsilon$-convergence time.
\end{theorem}

Theorems \ref{th:convrate} and \ref{th:preconvrate} suggest that
initializing with $k_0 = k^\beta$ leads to an improvement in
convergence time.  However, in our computational experience, we have
found that an initial condition of $k_0 = 0$ consistently results in
{\it faster} convergence than $k_0 = k^\beta$.  Hence, we suspect that
a convergence time bound of $O((\tau^\star / \epsilon)
\log(1/\epsilon))$ also holds for the case of $k_0 = 0$.  Proving this
remains an open issue.

\subsection{Adaptive Mixing Time Search}

The choice of $\beta$ is critical in that it determines both 
convergence time and ultimate accuracy. This raises the question of how 
to choose $\beta$ for a particular graph. The choices posited
in Theorems \ref{th:convrate} and \ref{th:preconvrate} 
require knowledge of $\tau^\star$, which may be both
difficult to compute and also requires knowledge of the graph
topology. This counteracts our purpose of developing a distributed
protocol.

In order to address this concern, consider
Algorithm~\ref{alg:varybeta}, which is designed for the case of $d > 2$. 
It uses a doubling sequence of guesses
$\tilde{\tau}$ for the Ces\'aro mixing time $\tau^\star$.  Each guess leads to
a choice of $\beta$ and a number of iterations $t^*$.  
Note that the algorithm takes $\epsilon > 0$ as input.

\begin{algorithm}
\caption{Synchronous consensus propagation with adaptive mixing time search.}
\label{alg:varybeta}
\begin{algorithmic}[1]
\STATE $K^{(0)} \leftarrow 0$, $\mu^{(0)} \leftarrow \hat{y}$, $t \leftarrow 0$
\FOR{$\ell=0$ to $\infty$}
\STATE $\tilde{\tau} \leftarrow 2^\ell$
\STATE Set $\beta$ and $t^*$ as indicated by Theorem \ref{th:convrate}, assuming $\tau^\star = \tilde{\tau}$
\FOR{$s=1$ to $t^*$}
\STATE $\mu^{(t)} \leftarrow \Gscr(\mu^{(t-1)},K^{(t-1)})$, $K^{(t)} \leftarrow \Fscr(K^{(t)})$
\STATE $t \leftarrow t+1$
\ENDFOR
\ENDFOR
\end{algorithmic}
\end{algorithm}

Consider applying this procedure to a $d$-regular graph with fixed $d
> 2$ but topology otherwise unspecified.  It follows from
Theorem~\ref{th:convrate} that this procedure has an
$\epsilon$-convergence time of $O((\tau^\star/\epsilon)
\log(\tau^\star/\epsilon))$.  An entirely analogous algorithm can be
designed for the case of $d = 2$.

We expect that many variations of this procedure can be made effective.
Asynchronous versions would involve each node adapting a local
estimate of the mixing time.

\section{Comparison with Pairwise Averaging}

Using the results of Section~\ref{se:convrate}, we can compare the
performance of consensus propagation to that of pairwise
averaging. Pairwise averaging is usually defined in an asynchronous
setting, but there is a synchronous counterpart which works as
follows.  Consider a doubly stochastic symmetric matrix
$P\in\R^{n\times n}$ such that $P_{ij}=0$ if $i\neq j$ and
$(i,j)\notin E$.  Evolve estimates according to $x^{(t)} = P
x^{(t-1)}$, initialized with $x^{(0)} = y$.  Here, at each time $t$, a
node $i$ is computing a new estimate $y^{(t)}_i$ which is an average
of the estimates at node $i$ and its neighbors during the previous
time-step. If the matrix $P$ is aperiodic and irreducible, then
$x^{(t)}=P^t y\tends \bar{y} \1$ as $t \uparrow \infty$.

In the case of a singly-connected graph, synchronous consensus
propagation converges exactly in a number of iterations equal to
the diameter of the graph. Moreover, when $\beta = \infty$, this
convergence is to the exact mean, as discussed
in Section~\ref{se:intuitive}. This is the best one can hope for under any
algorithm, since the diameter is the minimum amount of time required
for a message to travel between the two most distant nodes.  On the
other hand, for a fixed accuracy $\epsilon$, the worst-case number of
iterations required by synchronous pairwise averaging on a
singly-connected graph scales at least quadratically in the diameter
\cite{Boyd03}.

The rate of convergence of synchronous pairwise averaging is governed
by the relation $\| x^{(t)} - \bar{y} \1 \|_{2,n} \leq \lambda_2^t$,
where $\lambda_2$ is the second largest eigenvalue\footnote{Here, we
  take the standard approach of ignoring the smallest eigenvalue of
  $P$. We will assume that this eigenvalue is smaller than $\lambda_2$
  in magnitude. Note that a constant probability can be added to each
  self-loop of any particular matrix $P$ so that this is true.} 
of $P$.  Let $\tau_2 =
1/\log(1/\lambda_2)$, and call it the {\it mixing time} of $P$.  In
order to guarantee $\epsilon$-accuracy (independent of $y$), $t >
\tau_2 \log(1/\epsilon)$ suffices and $t = \Omega(\tau_2
\log(1/\epsilon))$ is required.

Consider $d$-regular graphs and fix a desired error tolerance
$\epsilon$.  The number of iterations required by consensus
propagation is $\Theta(\tau^\star \log \tau^\star)$, whereas that
required by pairwise averaging is $\Theta(\tau_2)$.  Both mixing times
depend on the size and topology of the graph.  $\tau_2$ is the mixing
time of a process on nodes that transitions along edges whereas
$\tau^\star$ is the mixing time of a process on directed edges that
transitions towards nodes.  An important distinction is that the
former process is allowed to ``backtrack'' where as the latter is not.
By this we mean that a sequence of states $(i,j,i)$ can be
observed in the vertex process, but the sequence $(\{i,j\}, \{j,i\})$
cannot be observed in the edge process.  As we will now illustrate
through an example, it is this difference that makes $\tau_2$ larger
than $\tau^\star$ and, therefore, pairwise averaging less efficient
than consensus propagation.

In the case of a cycle ($d=2$) with an even number of nodes $n$,
minimizing the mixing time over $P$ results in $\tau_2 = \Theta(n^2)$
\cite{Boyd04,Boyd05,Boyd05a}. For comparison, as demonstrated in the
following theorem (whose proof is deferred until the appendix),
$\tau^\star$ is linear in $n$.
\begin{theorem}\label{th:taucycle}
For the cycle with $n$ nodes, $\tau^\star \leq n/\sqrt{2}$.
\end{theorem}
Intuitively, the improvement in mixing time arises from the fact that
the edge process moves around the cycle in a single direction and
therefore travels distance $t$ in order $t$ iterations.  The vertex
process, on the other hand, is ``diffusive'' in nature. It randomly
transitions back and forth among adjacent nodes, and requires order
$t^2$ iterations to travel distance $t$. Non-diffusive methods
have previously been suggested in the design of efficient algorithms
for Markov chain sampling (see \cite{Diaconis00} and references
therein).

The cycle example demonstrates a $\Theta(n/\log n)$ advantage offered
by consensus propagation.  Comparisons of mixing times associated with
other graph topologies remains an issue for future analysis.  Let us
close by speculating on a uniform grid of $n$ nodes over the
$m$-dimensional unit torus.  Here, $n^{1/m}$ is an integer, and each
vertex has $2 m$ neighbors, each a distance $n^{-1/m}$ away.  With $P$
optimized, it can be shown that $\tau_2 = \Theta(n^{2/m})$
\cite{Roch04}.  We put forth a conjecture on $\tau^\star$.
\begin{conjecture}
For the $m$-dimensional torus with $n$ nodes,
$\tau^\star = \Theta(n^{(2m-1)/m^2})$.
\end{conjecture}

\section*{Acknowledgment}
\addcontentsline{toc}{section}{Acknowledgment}

The authors wish to thank Balaji Prabhakar and Ashish Goel for their
insights and comments. 

%\bibliography{IEEEabrv,average}

\begin{biography}{Ciamac C. Moallemi}
  Ciamac C. Moallemi is a PhD student in the Department of Electrical
  Engineering at Stanford University. He received SB degrees in
  Electrical Engineering and Computer Science and in Mathematics from the
  Massachusetts Institute of Technology (1996). He studied at the
  University of Cambridge as a British Marshall Scholar, where he
  earned a Certificate of Advanced Study in Mathematics, with
  distinction (1997).
\end{biography}

\begin{biography}{Benjamin Van Roy}
  Benjamin Van Roy is an Associate Professor of Management Science and
  Engineering, Electrical Engineering, and, by courtesy, Computer
  Science, at Stanford University, where he has been since 1998. His
  recent research interests include dynamic optimization, machine
  learning, economics, finance, and information technology. He
  received the SB (1993) in Computer Science and Engineering and the
  SM (1995) and PhD (1998) in Electrical Engineering and Computer
  Science, all from MIT. He is a member of INFORMS and IEEE. He serves
  on the editorial boards of Discrete Event Dynamic Systems, Machine
  Learning, Mathematics of Operations Research, and Operations
  Research.  He has been a recipient of Stanford's Tau Beta Pi Award
  for Excellence in Undergraduate Teaching, the NSF CAREER Award, and
  MIT's George M. Sprowls Dissertation Award.
\end{biography}

\appendices

\section{Proof of Theorem~\ref{th:conv}}

Given initial vectors $(\mu^{(0)},K^{(0)})$, and a sequence of communication sets
$\{ U_1, U_2,\ldots \}$, the consensus propagation algorithm evolves
parameter values over time according to
\begin{align}\label{eq:asyncupdate}
K_{ij}^{(t)} & = 
\begin{cases}
\Fscr_{ij}(K^{(t-1)}), & \text{if $\{i,j\}\in U_t$}, \\
K_{ij}^{(t-1)}, & \text{otherwise},
\end{cases}
\\
\mu_{ij}^{(t)} & = 
\begin{cases}
\Gscr_{i j}(\mu^{(t-1)},K^{(t-1)}), & \text{if $\{i,j\}\in U_t$}, \\
\mu_{ij}^{(t-1)}, & \text{otherwise},
\end{cases}
\end{align}
for times $t>0$.

In order to establish Theorem~\ref{th:conv}, we will first study
convergence of the inverse variance parameters $K^{(t)}$, and subsequently the
mean parameters $\mu^{(t)}$.

\subsection{Convergence of Inverse Variance Updates}

Our analysis of the convergence of the inverse variance parameters
follows the work in \cite{Rusmevichientong01}. We begin with a
fundamental lemma.
\begin{lemma}\label{le:fprop}
For each $\{i,j\}\in\vec{E}$, the following facts hold:
\begin{enumerate}
\item[(i)] The function $\Fscr_{ij}(\cdot)$ is continuous.
\item[(ii)] The function $\Fscr_{ij}(\cdot)$ is monotonic. That is, if $K
  \leq K'$, where the inequality is interpreted component-wise,
  then $\Fscr_{ij}(K) \leq \Fscr_{ij}(K')$.
\item[(iii)] If $K_{ij}'=\Fscr_{ij}(K)$, then $0 < K'_{ij} < \beta Q_{ij}$.
\item[(iv)] If $\alpha > 1$, then $\alpha \Fscr_{ij}(K) > \Fscr_{ij}
  (\alpha K)$.
\end{enumerate}
\end{lemma}
\begin{proof} Define the function $f: \R_+\rightarrow\R_+$  by
\[
f(x) = \cfrac{1}{\gamma + \cfrac{1}{1 + x}},
\]
where $\gamma > 0$. (i) follows from the fact that $f$ is
continuous. (ii) follows from the fact that $f(x)$ is strictly
increasing.  (iii) follows from the fact that $f(x)\in (0,1/\gamma)$
for all $x \geq 0$.  (iv) follows from the fact that $\alpha f(x) \geq
f(\alpha x)$.
\end{proof}

Now we consider the sequence of iterates $\{ K^{(0)},K^{(1)},\ldots \}$ which
evolve according to \eqref{eq:asyncupdate}.
\begin{lemma}
Let $K^{(0)}$ be such that $\Fscr_{ij}(K^{(0)}) \geq K^{(0)}$ for all
$\{i,j\}\in\vec{E}$ (for example, $K^{(0)} = 0$).  Then $K^{(t)}$ converges to a
vector $K^\beta$ such that $K^\beta = \Fscr(K^\beta)$.
\end{lemma}
\begin{proof}
Convergence follows from the fact that the iterates are component-wise
bounded and monotonic. The limit point must be a fixed point by
continuity.
\end{proof}

Given the existence of a single fixed point, we can establish that the
fixed point must be unique.
\begin{lemma}\label{le:varfixed}
The $\Fscr$ operator has a unique fixed point $K^\beta$.
\end{lemma}
\begin{proof}
  Denote $K^\beta$ to be the fixed point obtained by iterating with
  initial condition $K^{(0)}=0$, and let $K'$ be some other fixed
  point. It is clear that $K^{(0)} < K'$, thus, by monotonicity, we
  must have $K^\beta \leq K'$.  Define
\[
\gamma = \inf \left\{ \alpha \in [1,\infty) :\ K' \leq \alpha K^\beta \right\}.
\]
It is clear that $\gamma$ is well-defined since $0 < \{
K_{ij}^\beta,K'_{ij} \} < \beta Q_{ij}$. Also, we must have $\gamma >
1$, since $K^\beta\neq K'$. Then,
\[
K'_{ij} = \Fscr_{ij}(K') \leq \Fscr_{ij} (\gamma K^\beta) 
< \gamma \Fscr_{ij}(K^\beta) = \gamma K^\beta_{ij}.
\]
This contradicts the definition of $\gamma$. Hence, there is a unique
fixed point.
\end{proof}

\begin{lemma}\label{le:varconv}
Given an arbitrary initial condition $K^{(0)}\in\R_+^{|\vec{E}|}$,
\[
\lim_{t\tends\infty} K^{(t)} = K^\beta.
\]
\end{lemma}
\begin{proof}
If $0 \leq K^{(0)} \leq K^\beta$, the result holds by monotonicity. Assume
that $K^\beta \leq K^{(0)}$. 
\[
\gamma = \inf \left\{ \alpha \in [1,\infty) :\ K^{(0)} \leq \alpha K^\beta \right\}.
\]
Then,
\[
K^\beta_{ij} \leq \Fscr_{ij}(K^{(0)}) \leq \Fscr_{ij}(\gamma K^\beta)
\leq \gamma \Fscr_{ij}(K^\beta) = \gamma K^\beta_{ij}.
\]
Define a sequence $\{ \tilde{K}^{(t)} \}$ by
\[
\tilde{K}^{(0)} = \gamma K^\beta,
\]
and, for all $\{i,j\}\in\vec{E}$, $t > 0$,
\[
\tilde{K}_{ij}^{(t)} = 
\begin{cases}
\Fscr_{ij}(\tilde{K}^{(t-1)}), & \text{if $\{i,j\}\in U_t$}, \\
K_{ij}^{(t-1)}, & \text{otherwise}.
\end{cases}
\]
Since $\Fscr_{i j}(\tilde{K}^{(0)}) \leq \gamma \Fscr_{ij}(K^\beta) =
\tilde{K}_{ij}^{(0)}$, the sequence $\{ \tilde{K}^{(t)} \}$ is monotonically
decreasing and must have a limit which is a fixed point. Since the
fixed point is unique, we have $\tilde{K}^{(t)}\tends K^\beta$. But,
$K^\beta \leq K^{(0)} \leq \tilde{K}^{(0)}$. By monotonicity, we also have
$\tilde{K}^{(t)}\tends K^\beta$.

Now, consider the case of general $K^{(0)}$. Define $\underline{K}$ and
$\overline{K}$ such that $\underline{K} \leq K^{(0)} \leq \overline{K}$
and $\underline{K} \leq K^\beta \leq \overline{K}$. By the previous
two cases and monotonicity, we again have $K^{(t)}\tends K^\beta$.
\end{proof}

\subsection{Convergence of Mean Updates}

In this section, we will consider certain properties of the updates
for the mean parameters. Define the operator $\Gscr(\cdot,K)$ to
be the synchronous update of all components of the mean vector
according to
\[
\Gscr(\mu,K)_{ij} = \Gscr_{ij}(\mu,K),\quad\forall\ \{i,j\}\in\vec{E}.
\]

\begin{lemma}\label{le:contract}
  There exists $\alpha\in (0,1)$ so that 
\begin{enumerate}
\item[(i)] For all $\mu,\mu'\in\R^{\vec{E}}$,
\[
\| \Gscr(\mu,K^\beta)  - \Gscr(\mu',K^\beta) \|_\infty
< \alpha \| \mu - \mu' \|_\infty.
\]
\item[(ii)]
If $t$ is sufficiently large, for all $\mu,\mu'\in\R^{\vec{E}}$,
\[
\| \Gscr(\mu,K^{(t)})  - \Gscr(\mu',K^{(t)}) \|_\infty
< \alpha \| \mu - \mu' \|_\infty.
\]
\end{enumerate}
\end{lemma}
\begin{proof}
Set
\[
\bar{\alpha} = 
\max_{\substack{\{i,j\}\in\vec{E}\\u\in N(i)\setminus j}}
\frac{K^\beta_{ui}}{1 + \sum_{u'\in N(i)\setminus j} K^\beta_{u'i}}.
\]
Observing that $\bar{\alpha} < 1$, Part~(i) follows.

Define
\[
\bar{\alpha}_t = 
\max_{\substack{\{i,j\}\in\vec{E}\\u\in N(i)\setminus j}}
\frac{K^{(t)}_{ui}}{1 + \sum_{u'\in N(i)\setminus j} K^{(t)}_{u'i}}.
\]
Since $K^{(t)}\tends K^{\beta}$, by continuity
  $\bar{\alpha}_t\tends \bar{\alpha}<1$. Then, Part~(ii) follows.
\end{proof}

Lemma~\ref{le:contract} states that $\Gscr(\cdot,K^\beta)$ is a
maximum norm contraction. This leads to the following lemma.

\begin{lemma}\label{le:meanfixed}
The following hold:
\begin{enumerate}
\item[(i)] There is unique fixed point $\mu^\beta$ such that 
\[
\mu^\beta = \Gscr(\mu^\beta,K^\beta).
\]
\item[(ii)] There exists $T_1$ such if $t \geq T_1$, the operator
  $\Gscr(\cdot,K^{(t)})$ has a unique fixed point $\nu^{(t)}$. That
  is,
\[
\nu^{(t)} = \Gscr(\nu^{(t)},K^{(t)}).
\]
\item[(iii)] For any $\epsilon > 0$, there exists $T_2 \geq T_1$ so
  that if $t \geq T_2$,
\[
\| \nu^{(t)} - \mu^\beta \|_\infty < \epsilon.
\]
\end{enumerate}
\end{lemma}
\begin{proof}
  For Part~(i), since $\Gscr(\cdot,K^{\beta})$ is a maximum norm
  contraction, existence of a unique fixed point $\mu^\beta$ follows
  from, for example, Proposition~3.1.1 in
  \cite{BertsekasPDP}. Part~(ii) is established similarly.

  For Part~(iii), note for $t$ sufficiently large, the linear
  system of equations
\[
\nu = \Gscr(\nu,K^{(t)})
\]
over $\nu\in \R^{\vec{E}}$ is non-singular, by Part~(ii). Since
$K^{(t)}\tends K^\beta$, the coefficients of this system of equations
continuously converge to those of
\[
\nu = \Gscr(\nu,K^{\beta}).
\]
Then, we must have $\nu^{(t)}\tends \mu^\beta$.
\end{proof}

\subsection{Overall Convergence}

We are now ready to prove Theorem~\ref{th:conv}.

\renewcommand{\thetheorem}{\ref{th:conv}}
\begin{theorem}
  Assume that the communication sets $\{ U_t \}$ have the property
  that every directed edge $\{i,j\}\in \vec{E}$ appears in $U_t$ for
  infinitely many $t$. The following hold:
\begin{itemize}{\setlabelwidth{(iii)}}
\item[(i)] There are unique vectors $(\mu^\beta,K^\beta)$ such that
\[
K^\beta = \Fscr(K^\beta),
\quad\text{and}\quad
\mu^\beta = \Gscr(\mu^\beta,K^\beta).
\]
\item[(ii)] Independent of the initial condition $(\mu^{(0)},K^{(0)})$,
\[
\lim_{t\tends\infty} K^{(t)} = K^\beta,
\quad\text{and}\quad
\lim_{t\tends\infty} \mu^{(t)} = \mu^\beta.
\]
\item[(iii)] Given $(\mu^\beta,K^\beta)$, if
  $x^\beta=\Xscr(\mu^\beta,K^\beta)$, then $x^\beta$ is the mode of
  the distribution $p^\beta(\cdot)$.
\end{itemize}
\end{theorem}
\begin{proof}
  Existence and uniqueness of the fixed point $K^\beta$ and
  convergence of the vector $K^{(t)}$ to $K^\beta$ follow from
  Lemmas~\ref{le:varfixed} and \ref{le:varconv}, respectively.
  Existence and uniqueness of the fixed point $\mu^\beta$ follows from
  Lemma~\ref{le:meanfixed}.

  To establish the balance of Part~(ii), we need to show that
  $\mu^{(t)}\tends\mu^\beta$.  We will use a variant of the ``box
  condition'' argument of Proposition~6.2.1 in \cite{BertsekasPDP}.

Fix any $\epsilon>0$. By
Lemma~\ref{le:meanfixed}, pick $T_2$ so that if $t \geq T_2$, then
$\nu^{(t)}$ exists with $\nu^{(t)}=\Gscr(\nu^{(t)},K^{(t)})$ and
$\|\nu^{(t)}-\mu^\beta\|_\infty < \epsilon$. For $t > T_2$, if
$\{i,j\}\in U_t$,
\begin{equation}\label{eq:edgein}
\begin{split}
| \mu^{(t)}_{ij} - \mu^\beta_{ij} |
& 
\leq
| \mu^{(t)}_{ij} - \nu^{(t-1)}_{ij} |
+
|\nu^{(t-1)}_{ij} - \mu^\beta_{ij} |
\\
&
=
| \Gscr_{ij}(\mu^{(t-1)},K^{(t-1)}) - \Gscr_{ij}(\nu^{(t-1)},K^{(t-1)}) |
\\
&
\quad
+\:
\|\nu^{(t-1)} - \mu^\beta \|_\infty
\\
&
<
\alpha \| \mu^{(t-1)} - \nu^{(t-1)} \|_\infty
+
\|\nu^{(t-1)} - \mu^\beta \|_\infty
\\
&
\leq
\alpha \| \mu^{(t-1)} - \mu^\beta \|_\infty
+
(1 + \alpha) \|\nu^{(t-1)} - \mu^\beta \|_\infty
\\
&
\leq
\alpha \| \mu^{(t-1)} - \mu^\beta \|_\infty + (1+\alpha)\epsilon.
\end{split}
\end{equation}

For $k \geq 0$, define $\Ascr_k$ to be the set of vectors $\mu\in\R^{|\vec{E}|}$ such that
\[
\|\mu - \mu^\beta \|_\infty 
\leq 
\alpha^{k}
\|\mu^{(T_2)} - \mu^\beta \|_\infty 
+ (1+\alpha)\epsilon/(1-\alpha).
\]
We would like to show that for every $k \geq 0$, there is a time
$t_k$ such that $\mu^{(t)}\in\Ascr_k$, for all $t \geq t_k$. We
proceed by induction. 

When $k=0$, set $t_k=T_2$. Clearly $\mu^{(T_2)} \in \Ascr_0$. Assume
that $\mu^{(t-1)}\in\Ascr_0$, for some $t > T_2$. Then, if $\{i,j\}\in
U_{t}$, from \eqref{eq:edgein},
\[
\begin{split}
| \mu^{(t)}_{ij} - \mu^\beta_{ij} |
& <
\alpha \| \mu^{(t-1)} - \mu^\beta \|_\infty + (1+\alpha)\epsilon
\\
& <
\alpha \| \mu^{(T_2)} - \mu^\beta \|_\infty 
+ 
\frac{1+\alpha}{1 - \alpha}\alpha \epsilon
+
(1+\alpha)\epsilon
\\
& <
\| \mu^{(T_2)} - \mu^\beta \|_\infty 
+ 
\frac{1+\alpha}{1 - \alpha}\epsilon.
\end{split}
\]
If $\{i,j\}\notin U_{t}$,
\[
| \mu^{(t)}_{ij} - \mu^\beta_{ij} |
=
| \mu^{(t-1)}_{ij} - \mu^\beta_{ij} |
<
\| \mu^{(T_2)} - \mu^\beta \|_\infty + \frac{1+\alpha}{1-\alpha}\epsilon.
\]
Thus, $\mu^{(t)}\in \Ascr_0$. By induction, $\mu^{(t)}\in\Ascr_0$ for
all $t \geq T_2$.  

Now, assume that $t_{k-1}$ exists, for some $k-1 \geq 0$. Let $t >
t_{k-1}$ be some time such that $\{i,j\}\in U_t$. Then, by
\eqref{eq:edgein} and the fact that $\mu^{(t-1)}\in\Ascr_{k-1}$,
\[
\begin{split}
| \mu^{(t)}_{ij} - \mu^\beta_{ij} |
& <
\alpha \| \mu^{(t-1)} - \mu^\beta \|_\infty + (1+\alpha)\epsilon
\\
& <
\alpha^k \| \mu^{(T_2)} - \mu^\beta \|_\infty + \frac{1+\alpha}{1-\alpha}\epsilon.
\end{split}
\]
For each $\{i,j\}\in\vec{E}$, let $\tau^k_{ij} > t_{k-1}$ be the
earliest time after $t_{k-1}$ that $\{i,j\}\in U_{\tau^k_{ij}}$. If we
set $t_k$ to be the largest of these times, we have
$\mu^{(t)}\in\Ascr_k$, for all $t \geq t_k$.

We have established that
\[
\limsup_{t\tends\infty} 
\| \mu^{(t)} - \mu^\beta \|_\infty
\leq 
\alpha^k \| \mu^{(T_2)} - \mu^\beta \|_\infty + \frac{1+\alpha}{1-\alpha}\epsilon,
\]
for all $k \geq 0$. Taking a limit as $k\tends\infty$, we have
\[
\limsup_{t\tends\infty} 
\| \mu^{(t)} - \mu^\beta \|_\infty
\leq 
\frac{1+\alpha}{1-\alpha}\epsilon.
\]
Since $\epsilon$ was arbitrary, we have the convergence
$\mu^{(t)}\tends\mu^\beta$.

Part~(iii) follows from the fact that Gaussian belief propagation,
when it converges, computes exact means
\cite{Weiss01,Rusmevichientong01,Wainwright03}.
\end{proof}

\section{Proofs of Theorems~\ref{th:convrate} and \ref{th:preconvrate}}

In this section, we will prove Theorems~\ref{th:convrate} and
\ref{th:preconvrate}. We will start with some preliminary lemmas.

\subsection{Preliminary Lemmas}

The following lemma provides bounds on $k^\beta$ and $\gamma^\beta$ in
terms of $\beta$.
\begin{lemma}\label{le:kbeta}
If $d=2$,
\[
2\sqrt{\beta} - 1/2 < k^\beta < 2\sqrt{\beta},
\]
\[
\frac{1}{2\sqrt{\beta} + 1} < \gamma^\beta < \frac{1}{2\sqrt{\beta} + 1/2}.
\]
If $d>2$,
\[
\left(1 - \frac{1}{d-1}\right) \beta - \frac{1}{d-1} < k^\beta < \beta,
\]
\[
\frac{1}{1 + (d-1)\beta} 
< \gamma^\beta <
\frac{1}
{(d-2) \beta}.
\]
\end{lemma}
\begin{proof}
Starting with the fixed point equation \eqref{eq:kfixedpt}, some algebra
leads to
\[
\frac{d-1}{\beta} (k^\beta)^2 + (2 + \frac{1}{\beta} - d) k^\beta - 1 = 0.
\]
The quadratic formula gives us
\[
\begin{split}
k^\beta 
=
 \frac{\beta}{2} - \frac{\beta + 1}{2 (d-1)}
+ \sqrt{\left(\frac{\beta}{2} - \frac{\beta + 1}{2 (d-1)}\right)^2
+ 4 \frac{\beta}{d-1}},
\end{split}
\]
from which it is easy to derived the desired bounds.
\end{proof}

The following lemma offers useful expressions for the fixed point
$\mu^\beta$ and the mode $x^\beta$.

\begin{lemma}
\begin{subequations}\label{eq:muxseries}
\begin{align}
\mu^\beta & = \sum_{\tau=0}^\infty \gamma^\beta (1 - \gamma^\beta)^\tau \hat{P}^\tau \hat{y},
\\
x^\beta & = \frac{y}{1 + d k^\beta} 
+
\frac{d k^\beta}{1 + d k^\beta}
\sum_{\tau=0}^\infty \gamma^\beta (1 - \gamma^\beta)^\tau A
\hat{P}^\tau \hat{y}.
\end{align}
\end{subequations}
\end{lemma}
\begin{proof}
If we consider the algorithm when $k_0=k^\beta$, then $k_t=k^\beta$
and $\gamma_t=\gamma^\beta$ for all $t\geq 0$. Then, using
\eqref{eq:muiter} and induction, we have
\begin{align*}
\mu^{(t)} & = \sum_{\tau=0}^t \gamma^\beta (1 - \gamma^\beta)^\tau
\hat{P}^\tau \hat{y},
\\
x^{(t)} & = \frac{y}{1 + d k^\beta} 
+
\frac{d k^\beta}{1 + d k^\beta}
\sum_{\tau=0}^t \gamma^\beta (1 - \gamma^\beta)^\tau A
\hat{P}^\tau \hat{y}.
\end{align*}
The result follows from the fact that as $t\tends\infty$,
$\mu^{(t)}\tends\mu^\beta$ and $x^{(t)}\tends x^\beta$
(Theorem~\ref{th:conv}).
\end{proof}

The following lemma provides an estimate of the distance between fixed
points $\mu^\beta$ and $\mu^{\beta'}$ in terms of $|\gamma^\beta -
\gamma^{\beta'}|$.

\begin{lemma}\label{le:mubetabetap}
Given $0 \leq \beta' < \beta$, we have
\[
\| \mu^\beta - \mu^{\beta'} \|_{2,nd} 
\leq
\tau^\star 
(\gamma^{\beta'} - \gamma^{\beta})
(1 + 4/\gamma^{\beta}).
\]
\end{lemma}
\begin{proof}
Using \eqref{eq:muxseries},
\[
\begin{split}
\lefteqn{
\| \mu^\beta - \mu^{\beta'} \|_{2,nd} 
}
& 
\\
& =
\Biggl\|
\sum_{\tau = 0}^\infty 
\gamma^\beta (1 - \gamma^\beta)^\tau \hat{P}^\tau \hat{y}
-
\sum_{\tau = 0}^\infty 
\gamma^{\beta'} (1 - \gamma^{\beta'})^\tau \hat{P}^\tau \hat{y}
\Biggr\|_{2,nd}
\\
&
\leq
\Biggl\|
\sum_{\tau = 0}^\infty 
\Bigl(
\gamma^\beta (1 - \gamma^\beta)^\tau 
-
\gamma^{\beta'} (1 - \gamma^{\beta'})^\tau
\Bigr) \hat{P}^\tau 
\Biggr\|_{2,nd}.
\end{split}
\]
Since
\[
\sum_{\tau = 0}^\infty 
\gamma^\beta (1 - \gamma^\beta)^\tau 
-
\gamma^{\beta'} (1 - \gamma^{\beta'})^\tau 
=
0,
\]
we have
\[
\begin{split}
\lefteqn{
\| \mu^\beta - \mu^{\beta'} \|_{2,nd} 
}
&
\\
&
\leq
\Biggl\|
\sum_{\tau = 0}^\infty 
\Bigl(
\gamma^\beta (1 - \gamma^\beta)^\tau 
-
\gamma^{\beta'} (1 - \gamma^{\beta'})^\tau 
\Bigr)
(\hat{P}^\tau - \hat{P}^\star)
\Biggr\|_{2,nd}
\\
& =
\Biggl\|
\sum_{\tau = 0}^\infty 
\sum_{s=\tau}^\infty
\Bigl(
(\gamma^\beta)^2 (1 - \gamma^\beta)^s 
\\
&
\quad\quad\quad\quad\quad\quad
-\:
(\gamma^{\beta'})^2 (1 - \gamma^{\beta'})^s
\Bigr)
(\hat{P}^\tau - \hat{P}^\star)
\Biggr\|_{2,nd}
\\
& =
\Biggl\|
\sum_{s=0}^\infty
\Bigl(
(\gamma^\beta)^2 (1 - \gamma^\beta)^s
\\
&
\quad\quad\quad\quad\quad\quad
-\:
(\gamma^{\beta'})^2 (1 - \gamma^{\beta'})^s
\Bigr)
\sum_{\tau = 0}^s
(\hat{P}^\tau - \hat{P}^\star)
\Biggr\|_{2,nd}
\\
& \leq
\tau^\star
\sum_{s=0}^\infty
|
(\gamma^\beta)^2 (1 - \gamma^\beta)^s
-
(\gamma^{\beta'})^2 (1 - \gamma^{\beta'})^s
|.
\end{split}
\]
Hence, we wish to bound the sum
\[
\Delta = \sum_{s=0}^\infty
|
(\gamma^\beta)^2 (1 - \gamma^\beta)^s
-
(\gamma^{\beta'})^2 (1 - \gamma^{\beta'})^s
|.
\]

Set
\[
T = \left\lfloor 2 
\frac{\log \gamma^{\beta'} - \log \gamma^\beta}
{\log(1-\gamma^\beta) - \log(1-\gamma^{\beta'})}
\right\rfloor.
\]
Note that 
\begin{align*}
(\gamma^\beta)^2 (1 - \gamma^\beta)^s
& \leq
(\gamma^{\beta'})^2 (1 - \gamma^{\beta'})^s,
&
\text{if $s \leq T$,}
\\
(\gamma^\beta)^2 (1 - \gamma^\beta)^s
& \geq
(\gamma^{\beta'})^2 (1 - \gamma^{\beta'})^s,
&
\text{if $s > T$.}
\end{align*}
Holding $\gamma^\beta$ fixed, it is easy to verify that $T$ is
non-decreasing as $\gamma^{\beta'}\downarrow \gamma^\beta$.  Hence,
\begin{equation}\label{eq:Tbound}
\begin{split}
T
& \leq
2 
\frac{\log \gamma^{\beta'} - \log \gamma^\beta}
{\log(1-\gamma^\beta) - \log(1-\gamma^{\beta'})}
\\
& \leq
\lim_{\gamma^{\beta'}\downarrow\gamma^\beta}
2 
\frac{\log \gamma^{\beta'} - \log \gamma^\beta}
{\log(1-\gamma^\beta) - \log(1-\gamma^{\beta'})}
\\
& =
2(1-\gamma^\beta)/\gamma^\beta.
\end{split}
\end{equation}

Using the above results,
\[
\begin{split}
\Delta 
& = 
\sum_{s=0}^T
\left(
(\gamma^{\beta'})^2 (1 - \gamma^{\beta'})^s
-
(\gamma^\beta)^2 (1 - \gamma^\beta)^s
\right)
\\
&
\quad
+
\sum_{s=T+1}^\infty
\left(
(\gamma^\beta)^2 (1 - \gamma^\beta)^s
-
(\gamma^{\beta'})^2 (1 - \gamma^{\beta'})^s
\right)
\\
& =
\gamma^{\beta'} - \gamma^\beta 
- 2 \gamma^{\beta'} (1 - \gamma^{\beta'})^{T+1}
+ 2 \gamma^{\beta} (1 - \gamma^{\beta})^{T+1}
\\
& \leq
\gamma^{\beta'} - \gamma^\beta 
+ 2 \gamma^{\beta'} 
\left(
(1 - \gamma^{\beta})^{T+1}
- (1 - \gamma^{\beta'})^{T+1}
\right).
\end{split}
\]
Now, note that if $0 < a \leq b \leq 1$, for integer $\ell > 0$,
\[
\begin{split}
b^\ell - a^\ell 
& = b^\ell(1 - (a/b)^\ell) 
\\
& = b^\ell (1 - a/b) \sum_{i=0}^{\ell-1} (a/b)^i
\\
& \leq \ell b^{\ell-1} (b - a) 
\\
& \leq \ell (b-a).
\end{split}
\]
Applying this inequality and using \eqref{eq:Tbound}, we have
\[
\begin{split}
\Delta 
&
\leq
(\gamma^{\beta'} - \gamma^\beta)
\left(
1 + 2 (T + 1) \gamma^{\beta'} 
\right)
\\
&
\leq
(\gamma^{\beta'} - \gamma^\beta)
\left(
1 +  2 \gamma^{\beta'}(2 / \gamma^{\beta} - 1)
\right)
\\
&
\leq
(\gamma^{\beta'} - \gamma^\beta)
(
1 +  4 \gamma^{\beta'} / \gamma^{\beta}
)
\\
&
\leq
(\gamma^{\beta'} - \gamma^\beta)
(
1 +  4 / \gamma^{\beta}
),
\end{split}
\]
which completes the proof.
\end{proof}

The following lemma characterizes the rate at which $\gamma_t
\downarrow \gamma^\beta$.

\begin{lemma}\label{le:gammacont}
Assume that $\gamma^\beta \leq \gamma_0 \leq 1$. Then, $\{ \gamma_t
\}$ is a non-increasing sequence and
\[
| \gamma_t - \gamma^\beta | \leq 
\frac{(d-1)^t}{\left(1/\beta + \gamma^\beta + d - 1\right)^{2t}}.
\]
\end{lemma}
\begin{proof}
Define the function
\[
f(\gamma) = \frac{1}{1 + \frac{d-1}{1/\beta + \gamma}}.
\]
Note that, from the definition of $\gamma_t$ and \eqref{eq:ktevolve},
$\gamma_t=f(\gamma_{t-1})$. Further, from the definition of
$\gamma^\beta$ and \eqref{eq:kfixedpt}, it is clear that
$\gamma^\beta=f(\gamma^\beta)$. Since $k_0 \leq k^\beta$, then
$\gamma_0 \geq \gamma^\beta$, and since $k_t\uparrow k^\beta$ (from
Lemma~\ref{le:fprop}(ii)), $\gamma_t\downarrow \gamma^\beta$. Also, if
$\gamma\in [\gamma^\beta,1]$,
\[
 f'(\gamma)  = 
\frac{d-1}{\left(1/\beta + \gamma + d - 1\right)^2}
\leq  \frac{d-1}{\left(1/\beta + \gamma^\beta + d - 1\right)^2}.
\]
Then, by the Mean Value Theorem,
\[
\begin{split}
|\gamma_t - \gamma^\beta|
&
=
|f(\gamma_{t-1}) - f(\gamma^\beta)|
\\
&
\leq
\max_{\gamma\in[\gamma^\beta,1]}
|f'(\gamma)| |\gamma_{t-1} - \gamma^\beta|
\\
&
\leq
\frac{d-1}{\left(1/\beta + \gamma^\beta + d - 1\right)^2} 
|\gamma_{t-1} - \gamma^\beta|
\\
&
\leq \frac{(d-1)^t}{\left(1/\beta + \gamma^\beta + d - 1\right)^{2t}} 
|\gamma_{0} - \gamma^\beta|.
\end{split}
\]
\end{proof}

The following lemma establishes a bound on the distance between $x^{(t)}$
and $\bar{y}\1$ in terms of the distance between $\mu^{(t)}$ and
$\mu^\beta$.

\begin{lemma}\label{le:xtybar}
\[
\| x^{(t)} - \bar{y}\1 \|_{2,n}
\leq
\gamma_t + \gamma^\beta \tau^\star + \| \mu^{(t)} - \mu^\beta \|_{2,nd}.
\]
\end{lemma}
\begin{proof}
First, note that, using \eqref{eq:muxseries},
\begin{equation}\label{eq:mubetaybar}
\begin{split}
\| \mu^\beta - \hat{P}^\star \hat{y} \|_{2,nd} 
&
=
\left\|
\sum_{\tau = 0}^\infty 
\gamma^\beta (1 - \gamma^\beta)^\tau \hat{P}^\tau
-
\hat{P}^\star
\right\|_{2,nd}
\\
& =
\left\|
\sum_{\tau = 0}^\infty 
\gamma^\beta (1 - \gamma^\beta)^\tau 
(\hat{P}^\tau
-
\hat{P}^\star)
\right\|_{2,nd}
\\
& =
\left\|
\sum_{\tau = 0}^\infty 
(\gamma^\beta)^2 \sum_{s=\tau}^\infty (1 - \gamma^\beta)^s
(\hat{P}^\tau
-
\hat{P}^\star)
\right\|_{2,nd}
\\
& \leq
(\gamma^\beta)^2
\sum_{s = 0}^\infty 
 (1 - \gamma^\beta)^s
\left\|
\sum_{\tau=0}^s
(\hat{P}^\tau
-
\hat{P}^\star)
\right\|_{2,nd}
\\
& \leq
\gamma^\beta \tau^\star.
\end{split}
\end{equation}

Next, using Theorem~\ref{th:xstarconv}, Lemma~\ref{le:kbeta}, and
\eqref{eq:mubetaybar}, we have
\[
\begin{split}
\bar{y} \1 & = \lim_{\beta\tends \infty} x^\beta
\\
& = \lim_{\beta\tends\infty}
\frac{y}{1 + d k^\beta}
+ \frac{d k^\beta}{1 + d k^\beta} A \mu^\beta
\\
& =
\lim_{\beta\tends\infty}
A \mu^\beta
\\
& =
A \hat{P}^\star \hat{y}.
\end{split}
\]

Now,
\[
\begin{split}
\| x^{(t)} - \bar{y}\1 \|_{2,n}
& \leq
\frac{1}{1 + d k_t}\| y - \bar{y} \1 \|_{2,n}
\\
&
\quad
+\:
\frac{dk_t}{1 + dk_t}
\| A \mu^{(t)} - \bar{y} \1\|_{2,n}
\\
& \leq
\gamma_t
+
\| A \mu^{(t)} - \bar{y} \1\|_{2,n}
\\
& \leq
\gamma_t
+
\| A \mu^{(t)} - A \hat{P}^\star \hat{y} \|_{2,n}
\end{split}
\]
By examining the structure of $A$, it follows from the
Cauchy-Schwartz Inequality that
\[
\| A (\mu^{(t)} -  \hat{P}^\star \hat{y}) \|_{2,n}
\leq
\| \mu^{(t)} - \hat{P}^\star \hat{y} \|_{2,nd}.
\]
Thus, using \eqref{eq:mubetaybar}
\[
\begin{split}
\| x^{(t)} - \bar{y}\1 \|_{2,n}
&
\leq
\gamma_t
+
\| \mu^{(t)} - \hat{P}^\star \hat{y} \|_{2,nd}
\\
&
\leq
\gamma_t
+
\| \mu^\beta - \hat{P}^\star \hat{y} \|_{2,nd}
+
\| \mu^{(t)} - \mu^{\beta} \|_{2,nd}
\\
&
\leq
\gamma_t
+
\gamma^\beta \tau^\star
+
\| \mu^{(t)} - \mu^{\beta} \|_{2,nd}.
\end{split}
\]
\end{proof}

\subsection{Proof of Theorem~\ref{th:convrate}}

Theorem~\ref{th:convrate} follows immediately from the following lemma.

\begin{lemma}\label{le:convrate}
Fix $\epsilon>0$, and pick $\beta$ so that
\begin{align*}
\beta 
& \geq
\max\left\{(2(1+\tau^\star)/\epsilon - 1/2)^2/4, 9/16 \right\}, 
& \text{if $d=2$},
\\
\beta 
& \geq\max\left\{2(1+\tau^\star)/(\epsilon(d-2)), 3/(d-2) \right\},
& \text{if $d>2$}.
\end{align*}
Assume that $k_0 \leq k^\beta$. Define
\[
t^*
=
\left(1 + 2\sqrt{\beta}\right)
\log\left(
\frac{
2 + 9 \tau^\star \left(5+8\sqrt{\beta}\right)
\left(1/2+\sqrt{\beta}\right)
}
{\epsilon/2}
\right),
\]
if $d=2$, and
\[
t^* 
=
\left(1 + (d-1)\beta\right)
\log\left(
\frac{
2 + 4 \tau^\star \left(5+4(d-1)\beta\right)
}
{\epsilon/2}
\right),
\]
if $d>2$.
Then, $t^*$ is an $\epsilon$-convergence time.
\end{lemma}

\begin{proof}
Let $\beta_t$ be the value of $\beta$ implied by $k_t$, that is, the
unique value such that $k_t=k^{\beta_t}$.  Define
\[
\Delta_t = \| \mu^{(t)} - \mu^{\beta_t}\|_{2,nd}.
\]
Note that the matrix $\hat{P}$ is doubly stochastic and hence
non-expansive under the $\|\cdot\|_{2,nd}$ norm. Then, from
\eqref{eq:muiter} and the fact that $\mu^{\beta_t}$ is a fixed point,
\[
\begin{split}
\Delta_t 
& =
\|
\gamma_t \hat{y} +  (1-\gamma_t) \hat{P}\mu^{(t-1)} 
- 
\gamma_t \hat{y} -  (1-\gamma_t) \hat{P}\mu^{\beta_t} 
\|_{2,nd}
\\
& = \|(1-\gamma_t) \hat{P}(\mu^{(t-1)} - \mu^{\beta_t}) \|_{2,nd}
\\
& \leq (1-\gamma_t) \| \mu^{(t-1)} - \mu^{\beta_t} \|_{2,nd}
\\
& \leq (1 - \gamma^\beta) \| \mu^{(t-1)} - \mu^{\beta_t} \|_{2,nd}
\\
& \leq (1 - \gamma^\beta) 
\left(
\Delta_{t-1}
+
\| \mu^{\beta_{t-1}} - \mu^{\beta_t} \|_{2,nd}
\right).
\end{split}
\]
Now, using Lemmas~\ref{le:mubetabetap} and \ref{le:gammacont},
\begin{equation}\label{eq:dt}
\begin{split}
\Delta_t 
& \leq (1 - \gamma^\beta) 
\left(
\Delta_{t-1}
+
\tau^* 
(\gamma_{t-1} - \gamma_t)
(1 + 4/\gamma_t)
\right)
\\
& \leq (1 - \gamma^\beta) 
\left(
\Delta_{t-1}
+
\tau^* 
\left(\gamma_{t-1} - \gamma^{\beta}\right)
\left(1 + 4/\gamma^{\beta}\right)
\right)
\\
& \leq (1 - \gamma^\beta) 
\left(
\Delta_{t-1}
+
\tau^* 
\alpha^{t-1}
\left(1 + 4/\gamma^{\beta}\right)
\right).
\end{split}
\end{equation}
Here, we define
\[
\alpha
=
\begin{cases}
1/(\gamma^\beta + 1)^2, & \text{if $d=2$}, \\
1/(d-1), & \text{if $d>2$}.
\end{cases}
\]

We would like to ensure that $\alpha < 1 - \gamma^\beta$. For $d=2$,
some algebra reveals that this is is true when $0 < \gamma^\beta <
(\sqrt{5}-1)/2$. By the fact that $\beta \geq 9/16$ and
Lemma~\ref{le:kbeta}, we have
\[
0 < \gamma^\beta < \frac{1}{2\sqrt{\beta} + 1} \leq 2/5 
< \frac{\sqrt{5}-1}{2}.
\]
For $d>2$, using the fact that $\beta\geq 3/(d-2)$ and Lemma~\ref{le:kbeta},
\begin{equation}\label{eq:d3ab}
\begin{split}
0
& <
\frac{\alpha}{1-\gamma^\beta}
<
\frac{(d-2)\beta}{(d-1)((d-2)\beta - 1)} 
\\
& < \frac{3}{2(d-1)} 
\leq 3/4 < 1.
\end{split}
\end{equation}

By induction using \eqref{eq:dt}, we have
\[
\begin{split}
\Delta_t 
&
\leq
(1 - \gamma^\beta)^t 
+
\tau^* 
\left(1 + 4/\gamma^{\beta}\right)
\sum_{s=0}^{t-1}
(1-\gamma^\beta)^{t-s} \alpha^s
\\
&
\leq
(1 - \gamma^\beta)^t 
\left(
1
+
\tau^* 
\frac{1 + 4/\gamma^{\beta}}{1 - \alpha/(1-\gamma^\beta)}
\right).
\end{split}
\]
Now, notice that using the above results and
Lemmas~\ref{le:mubetabetap}, \ref{le:gammacont}, and \ref{le:xtybar},
\[
\begin{split}
\lefteqn{
\| x^{(t)} - \bar{y} \1 \|_{2,n}
} &
\\
& \leq
\gamma_t + \gamma^\beta \tau^\star + \| \mu^{(t)} - \mu^\beta \|_{2,nd}
\\
& \leq
\gamma_t + \gamma^\beta \tau^\star + \Delta_t
 + \| \mu^{\beta_t} - \mu^\beta \|_{2,nd}
\\
&
\leq
\gamma^\beta (1 + \tau^\star)
+
(\gamma_t - \gamma^\beta)
+
\Delta_t
\\
&
\quad
+\:
\tau^\star
(\gamma_t - \gamma^\beta)
(1 + 4/\gamma^\beta)
\\
&
\leq
\gamma^\beta (1 + \tau^\star)
+
\alpha^t
\\
&
\quad
+\:
(1 - \gamma^\beta)^t 
\left(
1
+
\tau^* 
\frac{1 + 4/\gamma^{\beta}}{1 - \alpha/(1-\gamma^\beta)}
\right)
\\
&
\quad
+\:
\tau^\star
\alpha^t 
(1 + 4/\gamma^\beta)
\\
&
\leq
(1 - \gamma^\beta)^t 
\left(
2
+
\tau^* 
\left(1 + 4/\gamma^{\beta}\right)
\left(
1+
\frac{1}{1 - \alpha/(1-\gamma^\beta)}
\right)
\right)
\\
&
\quad
+\:
\gamma^\beta(1 + \tau^\star).
\end{split}
\]

When $d=2$, using Lemma~\ref{le:kbeta} and the fact that $\beta \geq
(2(1+\tau^\star)/\epsilon - 1/2)^2/4$, we have
\[
(1 + \tau^\star) \gamma^\beta 
< 
\frac{1+\tau^\star}{2\sqrt{\beta} + 1/2}
\leq \epsilon/2.
\]
Similarly, when $d>2$, since $\beta \geq 2(1+\tau^\star)/(\epsilon(d-2))$,
\[
(1 + \tau^\star) \gamma^\beta 
< 
\frac{1+\tau^\star}{(d-2)\beta}
\leq \epsilon/2.
\]

Thus, we will have $\|x^{(t)}-\bar{y}\1\|_{2,n}\leq \epsilon$ if
\begin{multline}
(1 - \gamma^\beta)^t 
\left(
2
+
\tau^* 
\left(1 + 4/\gamma^{\beta}\right)
\left(
1+
\frac{1}{1 - \alpha/(1-\gamma^\beta)}
\right)
\right)
\\
\leq \epsilon/2.
\end{multline}
This will be true when 
\begin{equation}\label{eq:tmin}
t
\geq
\frac{1}{\gamma^\beta}
\log\left(
\frac{
2
+
\tau^* 
\left(1 + 4/\gamma^{\beta}\right)
\left(
1+
\frac{1}{1 - \alpha/(1-\gamma^\beta)}
\right)
}
{\epsilon/2}
\right).
\end{equation}
(We have used the fact that $\log(1-\gamma^\beta) \leq
-\gamma^\beta$.)  To complete the theorem, it suffices to show that
$t^*$ is an upper bound to the right hand side of \eqref{eq:tmin}.

Consider the $d=2$ case.  From Lemma~\ref{le:kbeta}, it follows that
\[
1/\gamma^\beta < 1 + 2\sqrt{\beta},
\]
\[
1+4/\gamma^\beta < 5 + 8\sqrt{\beta}.
\]
Finally,
\[
\begin{split}
\frac{1}{1-\alpha/(1-\gamma^\beta)}
& =
\frac{1}{1-\frac{1}{(1+\gamma^\beta)^2 (1-\gamma^\beta)}}
\\
& =
\frac{1}{\gamma^\beta}
\frac
{(1+\gamma^\beta)^2 (1-\gamma^\beta)}
{1 - \gamma^\beta - (\gamma^\beta)^2}
\\
& =
\frac{h(\gamma^\beta)}{\gamma^\beta}.
\end{split}
\]
Since $\beta \geq 9/16$, from Lemma~\ref{le:kbeta},
$\gamma^\beta\in(0,1/2)$. It is easy to verify that for such
$\gamma^\beta$, the rational function $h(\gamma^\beta)$ satisfies
$h(\gamma^\beta) < h(1/2) = 9/2$. Thus,
\[
\frac{1}{1-\alpha/(1-\gamma^\beta)}
<
\frac{9}{2\gamma^\beta}
<
9/2 + 9\sqrt{\beta}.
\]

For the $d>2$ case, from Lemma~\ref{le:kbeta}, it follows that
\[
1/\gamma^\beta < 1 + (d-1)\beta,
\]
\[
1+4/\gamma^\beta \leq
5 + 4 (d-1) \beta.
\]
Finally, using \eqref{eq:d3ab}
\[
\frac{1}{1-\alpha/(1-\gamma^\beta)}
<
\frac{1}{1-3/4} = 4.
\]
\end{proof}

\subsection{Proof of Theorem~\ref{th:preconvrate}}

Theorem~\ref{th:preconvrate} follows immediately from the following lemma.

\begin{lemma}\label{le:preconvrate}
Fix $\epsilon>0$, and pick $\beta$ so that
\begin{align*}
\beta 
& \geq
(2(1+\tau^\star)/\epsilon - 1/2)^2/4,
& \text{if $d=2$},
\\
\beta 
& \geq
2(1+\tau^\star)/(\epsilon(d-2)),
& \text{if $d>2$}.
\end{align*}
Assume that $k_0 = k^\beta$, and define
\[
t^*
=
\begin{cases}
\left(1 + 2\sqrt{\beta}\right)
\log(2/\epsilon),
&
\text{if $d=2$,}
\\
\left(1 + (d-1)\beta\right)
\log(2/\epsilon),
&
\text{if $d>2$.}
\end{cases}
\]
Then, $t^*$ is an $\epsilon$-convergence time.
\end{lemma}

\begin{proof}
Note that in this case, we have $k_t=k^\beta$ and
$\gamma_t=\gamma^\beta$, for all $t\geq 0$. We will follow the
same strategy as the proof of Lemma~\ref{le:convrate}. Define
\[
\Delta_t = \| \mu^{(t)} - \mu^{\beta}\|_{2,nd}.
\]
Note that the matrix $\hat{P}$ is doubly stochastic and hence
non-expansive under the $\|\cdot\|_{2,nd}$ norm. Then, from
\eqref{eq:muiter} and the fact that $\mu^{\beta_t}$ is a fixed point,
\[
\begin{split}
\Delta_t 
& =
\|
\gamma^\beta \hat{y} +  (1-\gamma^\beta) \hat{P}\mu^{(t-1)} 
- 
\gamma^\beta \hat{y} -  (1-\gamma^\beta) \hat{P}\mu^{\beta} 
\|_{2,nd}
\\
& = \|(1-\gamma^\beta) \hat{P}(\mu^{(t-1)} - \mu^{\beta}) \|_{2,nd}
\\
& \leq (1-\gamma^\beta) \| \mu^{(t-1)} - \mu^{\beta} \|_{2,nd}
\\
& = (1 - \gamma^\beta) \Delta_{t-1}
\\
& \leq (1 - \gamma^\beta)^t,
\end{split}
\]
where the last step follows by induction.

Now, notice that, using the result and Lemmas~\ref{le:xtybar},
\[
\begin{split}
\| x^{(t)} - \bar{y} \1 \|_{2,n}
& \leq
\gamma^\beta(1 + \tau^\star) + \Delta_t
\\
& \leq
\gamma^\beta(1 + \tau^\star) + (1-\gamma^\beta)^t.
\end{split}
\]

When $d=2$, using Lemma~\ref{le:kbeta} and the fact that $\beta \geq
(2(1+\tau^\star)/\epsilon - 1/2)^2/4$, we have
\[
(1 + \tau^\star) \gamma^\beta 
< 
\frac{1+\tau^\star}{2\sqrt{\beta} + 1/2}
\leq \epsilon/2.
\]
Similarly, when $d>2$, since $\beta \geq 2(1+\tau^\star)/(\epsilon(d-2))$,
\[
(1 + \tau^\star) \gamma^\beta 
< 
\frac{1+\tau^\star}{(d-2)\beta}
\leq \epsilon/2.
\]

Thus, we will have $\|x^{(t)}-\bar{y}\1\|_{2,n}\leq \epsilon$ if
\[
(1 - \gamma^\beta)^t 
\leq \epsilon/2.
\]
This will be true when 
\begin{equation}\label{eq:tmin2}
t
\geq
\frac{1}{\gamma^\beta}
\log(2/\epsilon).
\end{equation}
(We have used the fact that $\log(1-\gamma^\beta) \leq
-\gamma^\beta$.)  To complete the theorem, it suffices to show that
$t^*$ is an upper bound to the right hand side of \eqref{eq:tmin2}.

Consider the $d=2$ case.  From Lemma~\ref{le:kbeta}, it follows that
\[
1/\gamma^\beta < 1 + 2\sqrt{\beta}.
\]

For the $d>2$ case, from Lemma~\ref{le:kbeta}, it follows that
\[
1/\gamma^\beta < 1 + (d-1)\beta.
\]
\end{proof}

\section{Proof of Theorem~\ref{th:taucycle}}

\renewcommand{\thetheorem}{\ref{th:taucycle}}
\begin{theorem}
For the cycle with $n$ nodes, $\tau^\star \leq n/\sqrt{2}$.
\end{theorem}
\begin{proof}
Let $e^{ij} \in \R^{2n}$ be the vector with $\{i,j\}$th component equal to
$1$ and each other component equal to $0$.  It is easy to see that for
any $\{i,j\} \in \vec{E}$,
\[
\begin{split}
\sup_t \left\|\sum_{\tau=0}^t (\hat{P}^\tau - \hat{P}^\star) e^{ij}\right\|_{2,2n}^2
&
= \left\|\sum_{\tau=0}^{\lfloor n/2 \rfloor} (\hat{P}^\tau - \hat{P}^\star)
e^{ij}\right\|_{2,2n}^2 
\\
& \leq \frac{1}{2 \sqrt{2}}.
\end{split}
\]
We then have
\[
\begin{split}
\tau^\star
& = 
\sup_{t,\mu}
\frac{
\left\|\sum_{\tau = 0}^t (\hat{P}^\tau - \hat{P}^\star) \mu\right\|_{2,2n}}
{\|\mu\|_{2,2n}} 
\\
&=
\sup_{t,\mu}
\frac{\left\|\sum_{\tau = 0}^t (\hat{P}^\tau - \hat{P}^\star)
\sum_{\{i,j\}} \mu_{ij} e^{ij}\right\|_{2,2n}}
{\left\|\sum_{\{i,j\}} \mu_{ij} e^{ij}\right\|_{2,2n}} \\
&\leq
\sup_{t,\mu}
\frac{
\sum_{\{i,j\}}
\mu_{ij} \left\|\sum_{\tau = 0}^t (\hat{P}^\tau - \hat{P}^\star)
e^{ij} \right\|_{2,2n}}
{\left\|\sum_{\{i,j\}} \mu_{ij} e^{ij}\right\|_{2,2n}} \\
& \leq
\sup_{\mu} \frac{\sum_{\{i,j\}} \mu_{ij}}
{2 \sqrt{2} \|\sum_{\{i,j\}} \mu_{ij} e^{ij}\|_{2,2n}} \\
& = 
\sup_{\mu} \frac{\sum_{\{i,j\}} \mu_{ij}}
{2 \sqrt{2} \sqrt{\sum_{\{i,j\}} \mu_{ij}^2 / 2n}} \\
& \leq
\sup_{\mu} \frac{\sum_{\{i,j\}} \mu_{ij}}
{2 \sqrt{2} \sum_{\{i,j\}} |\mu_{ij}| / 2n} \\
& \leq
\frac{n}{\sqrt{2}}.
\end{split}
\]
\end{proof}

\end{document}